\documentclass[onecolumn,amsmath,amssymb]{qm2008_abs}
\usepackage{graphicx}% Include figure files
\usepackage{dcolumn}% Align table columns on decimal point
\usepackage{bm}% bold math
\begin{document}

\def\simge{\mathrel{%
   \rlap{\raise 0.4ex \hbox{$>$}}{\lower 0.65ex \hbox{$\sim$}}}}
\def\simle{\mathrel{
   \rlap{\raise 0.4ex \hbox{$<$}}{\lower 0.65ex \hbox{$\sim$}}}}

\title{{\Large An analytic study towards instabilities of the glasma}}

\bigskip
\bigskip
\author{\large H. Fujii$^a$ and K. Itakura$^b$}
%\email{hfujii@phys.c.u-tokyo.ac.jp}
%\email{kazunori.itakura@kek.jp}
\affiliation{$^a$ Institute of Physics, University of Tokyo,
                  Komaba, Tokyo 153-8902, Japan\\
             $^b$ High Energy Accelerator Research Organization (KEK), 
                  %Oho 1-1, 
             Tsukuba, Ibaraki 305-0801, Japan}
\bigskip
\bigskip

\begin{abstract}
\leftskip1.0cm
\rightskip1.0cm
Strong longitudinal color flux fields will
be created in the initial stage of high-energy nuclear collisions.
We investigate analytically time evolution of such boost-invariant 
color fields from Abelian-like initial conditions,
and next examine stability 
of the boost-invariant configurations against rapidity
dependent fluctuations\cite{FujiiItakura}.
We find that the magnetic background field has an instability induced
by the lowest Landau level whose amplitude grows exponentially.
For the electric background field there is no apparent instability
although pair creations due to the Schwinger mechanism should be
involved.
\end{abstract}

\maketitle

\section{Introduction}

In the high-energy limit of nucleus-nucleus collisions, the relevant
degrees of freedom of the incident nucleus are the small-$x$ partons
described in the framework of the Color Glass Condensate
(CGC)\cite{CGC} with the typical momentum scale, 
or {\it  saturation scale} $Q_s(x)$.
On the other hand, success of hydrodynamic models in the
data analyses at RHIC seems to imply the formation of
quark-gluon plasma (QGP) in local equilibrium in a very early stage 
($\tau <$ 1 fm/$c$) of the event.
Then it must be a decisive challenge to understand the physical
mechanism for such a rapid equilibration from the CGC initial condition.

Plasma instabilities\cite{Plasma_instability}, especially the Weibel type, are intensively
investigated so far, where one analyzes coupled equations of
motion for the gauge fields and the hard particles with anisotropic
momentum distribution.
Despite that many intriguing phenomena have been reported, the 
kinetic description for particles with momentum $p_\perp \simeq Q_s$
should be applicable  only after particle formation $\tau \simge 1/Q_s$.
In the very early stage, $\tau \simle 1/Q_s$, 
the system is so dense that the field description may be
more appropriate. 
Here we shall investigate the instability problem 
in the classical Yang-Mills (YM) equations.

The dense pre-equilibrium system appearing between the first impact
and the equilibrated QGP was recently named
{\it Glasma}\cite{Lappi-McLerran}.
The glasma, produced from the CGC initial condition, is still
well-described by strong coherent YM field and can be treated
as a weakly coupled system. 
Reflecting the CGC initial condition,
this coherent field is boost-invariant in the
high energy limit,
and thus the
{\it time-evolution} towards a thermalized QGP should involve
the process exciting rapidity-dependent degrees of freedom.
If rapidity-dependent unstable modes exist in the glasma,
they will drive the system into an isotropic/thermalized state 
more efficiently.
In fact, numerical simulations for the glasma have shown
instability against the rapidity-dependent perturbations\cite{Raju}.
Here we present an analytic attempt for 
understanding the dynamics of the glasma instability\cite{FujiiItakura}.

\begin{figure}[t]
\begin{center}
\includegraphics[scale=0.4]{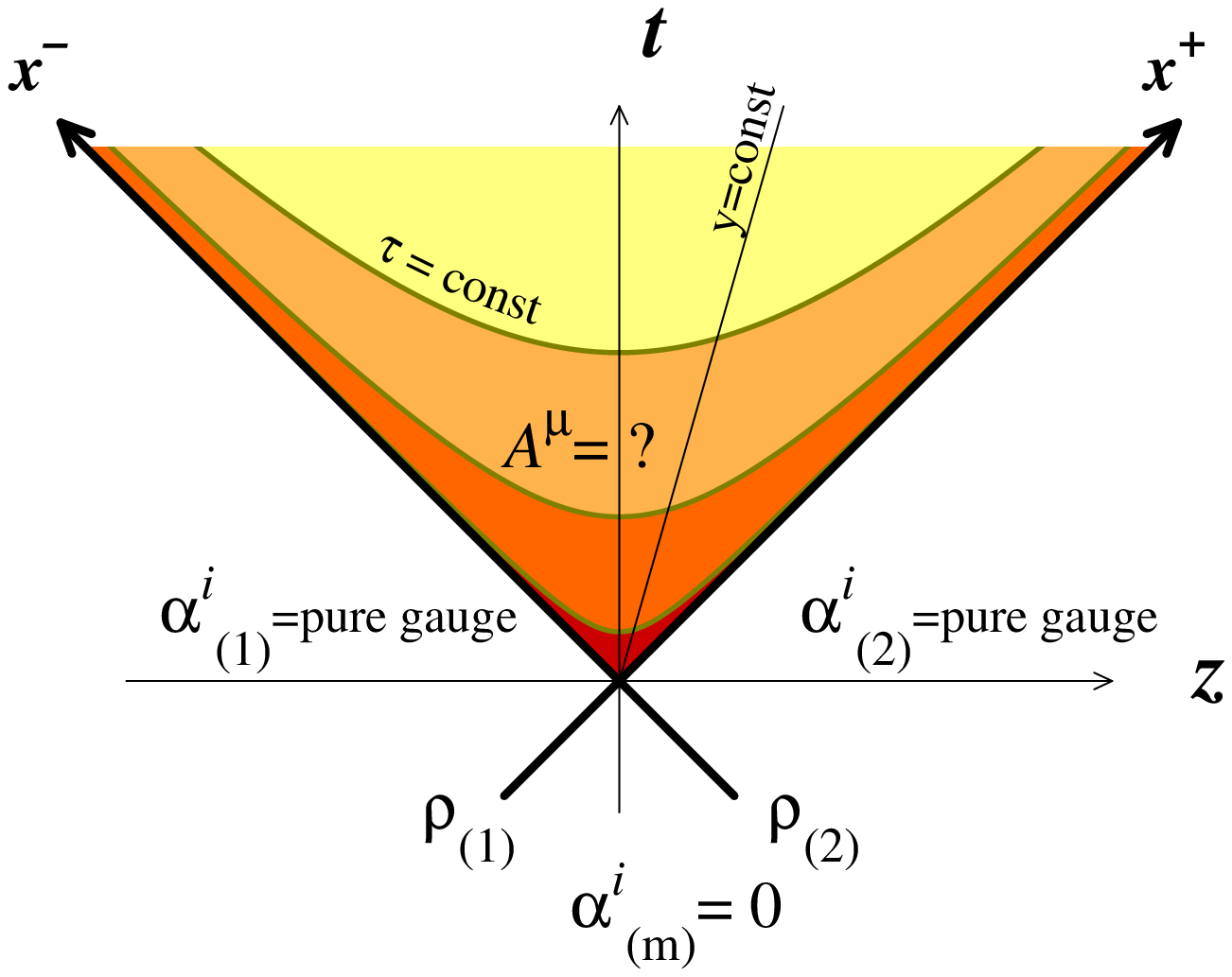}
\hfil
\includegraphics[scale=0.4]{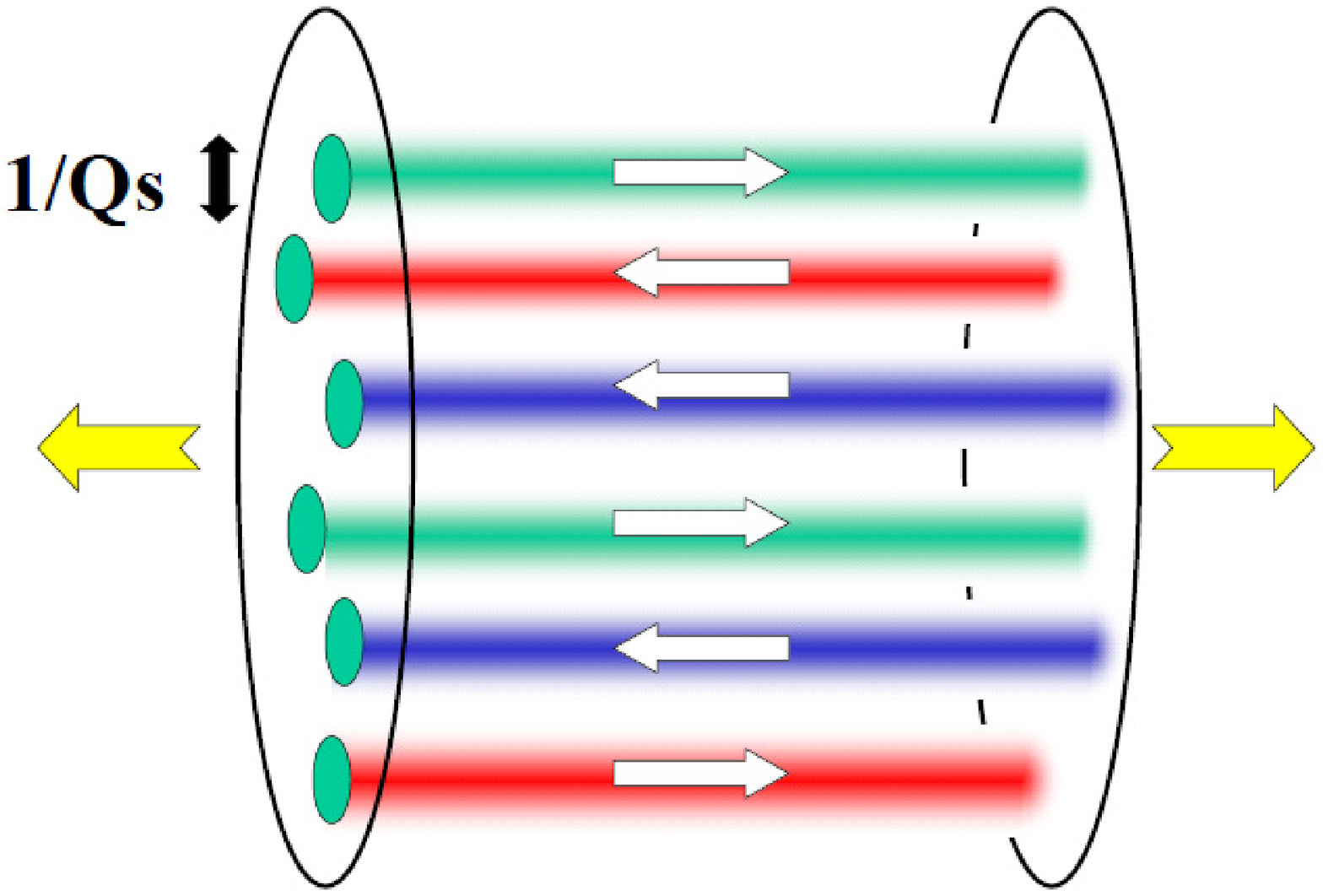}
\caption{Event setup (left) and 
schematic picture (right) of the color electric and magnetic flux
  tubes with transverse size 1/$Q_s$, created between the two
 Lorentz-contracted nuclei just after the collision.}
\end{center}
\end{figure}

\section{Expanding color flux tube}

In the CGC picture for the nuclear collisions,
the random color sources are set on the light-cone,
$x^\pm=(t\pm z)/\sqrt{2}$, and cross each other at the origin (Fig.~1).
The two sources are
accompanied with pure gauge fields $\alpha_{1,2}^{i}(x_\perp)$
 ($i=x, y$) behind,
which uniquely fix the initial condition for the gauge field in the
forward light-cone as
($\tau=\sqrt{2x^+ x^-}$ and $\eta= \frac12 \ln ({x^+}/{x^-})$)%
\cite{Glasma_basic}
\begin{eqnarray}
{\mathcal A}_\eta
&=&
-\tau^2 \alpha(\tau=0,x_\perp)
=\tau^2 \frac{ig}{2}[\alpha_1^i(x_\perp), \alpha_2^i(x_\perp)]\, , \\
{\mathcal A}_i 
&=&
\alpha^i (\tau=0,x_\perp)
= \alpha_1^i(x_\perp)+ \alpha_2^i(x_\perp)\, .
\end{eqnarray}
Here we use the Fock-Schwinger gauge, ${\mathcal A}_\tau =0$.
The corresponding initial field strengths are
\begin{eqnarray}
E^z\vert_{\tau=0^+} = -ig [\alpha_1^i, \alpha_2^i]\, , \qquad
B^z\vert_{\tau=0^+} = ig \epsilon_{ij}[\alpha_1^i, \alpha_2^j]\, .
\end{eqnarray}
with all transverse components vanishing. 
Note that this setup is perfectly longitudinal-boost invariant
and has no $\eta$ dependence while 
the transverse coherence length of the fields should be of order $1/Q_s$, 
reflecting the CGC structures of the
incident nuclei.
Thus we have the picture of multiple tubes of the boost-invariant 
color flux of size $1/Q_s$ distributed in the transverse plane.

One can compute the gluon production,
in principle, by solving the source-free YM equations in
the forward light-cone from this initial condition.
Event averaging corresponds to averaging over 
the initial color source distribution in the CGC framework.
Note, however, that we {\it don't} take this averaging,
but presume that thermalization should occur in each event.
In particular, we study here
the time-evolution of a single isolated magnetic/electric color
flux tube.

\vspace*{1mm}
{\it Magnetic flux tube} : 
A magnetic flux tube is the unique object in the glasma configuration.
This can be realized with
\begin{equation}
\alpha(\tau,x_\perp) = 0,\qquad \alpha_i(\tau, x_\perp) \neq 0\, . 
\end{equation}
For simplicity, we assume that the color direction of $\alpha_i$
and $B^z$ are the same and constant, and then
one can ignore the commutators.
The Gauss constraint is trivially
satisfied in this case, and
the YM equations reduce to the
Bessel equation\cite{FujiiItakura}.
For definiteness, we take the Gaussian profile in the transverse
plane for the initial condition:
\begin{equation}
\tilde \alpha_i^{\rm init}(k_\perp)\propto -i
\frac{\epsilon_{ij}k_j}{k_\perp^2}\, {\rm e}^{-\frac{k_\perp^2}{4Q_s^2}}\, .
\end{equation}
The solution is written in terms of the modified Bessel functions
$I_{0,1}(z)$:
\begin{eqnarray}
B^z(\tau \ge 0,r)
&=&
B_0 \, {\rm e}^{-Q_s^2r^2} {\rm e}^{-Q_s^2\tau^2 } I_0(2Q_s^2 r \tau )\, ,
\\
E^T(\tau \ge 0,r)
&=&
B_0\, {\rm e}^{-Q_s^2r^2} {\rm e}^{-Q_s^2\tau^2 }\,
I_1(2Q_s^2 r \tau )\, .
\end{eqnarray}
where $E^T\equiv \sqrt{(E^x)^2+(E^y)^2}$ and $r=|x_\perp|$.
This describes the outward expansion of the flux tube as seen in Fig.~2
(left and middle).
In the right panel of Fig.~2, we show the energy densities of the flux tube
integrated over the transverse plane, $\overline{(B^z)^2}$ and 
$\overline{(E^T)^2}$ as a function of time $\tau$. 
Remarkably, this time dependence is quite similar to the
numerical results in \cite{Lappi-McLerran}.

\begin{figure}[t]
\includegraphics[scale=0.6]{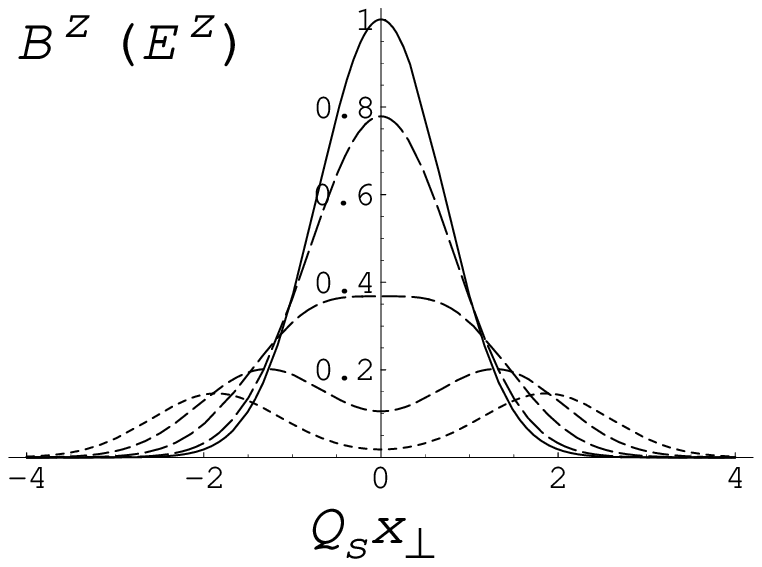}
\hfil
\includegraphics[scale=0.6]{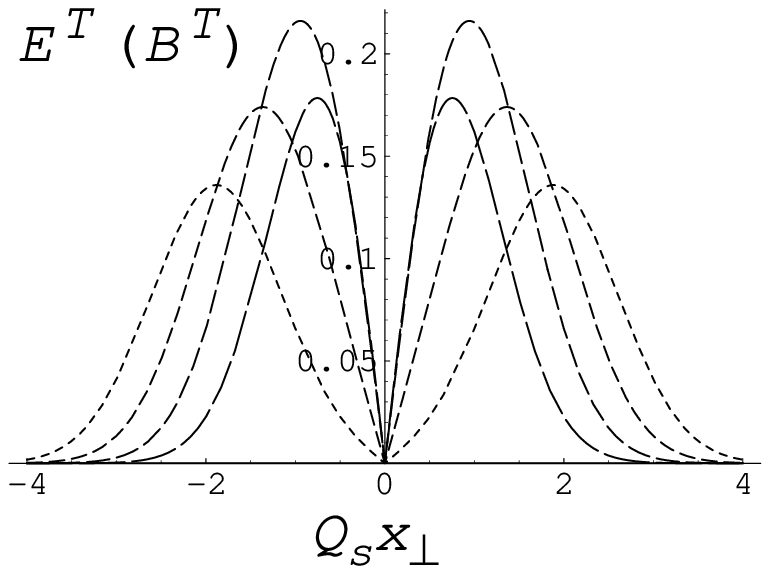}
\hfil
\includegraphics[scale=0.6]{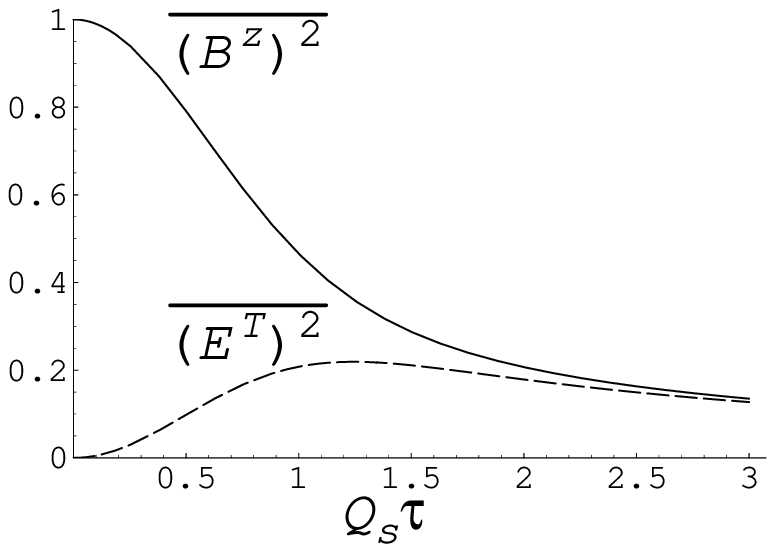} 
\caption{Spatial profile of the longitudinal magnetic field 
$B^z$ (left) and the transverse electric field
$E^T$ (middle),
and time dependence of the averaged field strengths,
$\overline{(B^z)^2}$ and $\overline{(E^T)^2}$  (right)
for a single flux tube.
The spatial profiles are plotted
at five different times $Q_s \tau  = 0 $ (solid),
\ 0.5 (longest dash),\ 1.0,\ 1.5,\ 2.0 (shortest dash)
with setting $B_0=1$. 
The initial value of $\overline{(B^z)^2}(\tau)$
is also set to 1.
Notice that $E^T=0$ at $\tau=0$. 
These plots are true for the electric flux tube after exchanging
$E$ and $B$.}
\end{figure}

\vspace*{1mm}
{\it Electric flux tube} :  
In contrast, an electric flux tube is realized with
\begin{equation}
\alpha(\tau, x_\perp)\neq 0,\qquad \alpha_i(\tau, x_\perp)=0.
\end{equation}
We assume again that the color orientations of $\alpha$ and $E^z$
are the same and constant. Then the YM equation reduces to
the Bessel equation. With the Gaussian initial condition
the following results are completely dual to the magnetic case by
exchanging $E$ and $B$.

\section{Instabilities of the glasma}

The outward expansion of the boost-invariant flux tubes never
achieves the isotropization of the system, because its longitudinal
pressure is never positive
($\propto \overline{(B^z)^2} -\overline{(E^T)^2}$ for the magnetic
case).
This is where the rapidity-dependent fluctuations play the leading part.

We introduce the small fluctuations $a_{i, \eta}$ as
\begin{eqnarray}
A_i = {\mathcal A}_i(\tau,x_\perp) +a_i(\tau,\eta,x_\perp)
\, ,\quad A_\eta = {\mathcal A}_\eta(\tau,x_\perp) + a_\eta(\tau,\eta,x_\perp)\, ,
\end{eqnarray}
where 
${\mathcal A}_i$ and ${\mathcal A}_\eta$ are boost-invariant 
background fields in the previous section.
We perform stability analysis of this system
against the perturbations $a_{i,\eta}$%
\cite{Chang-Weiss}.
Note that coupling between
${\mathcal A}_{i,\eta}$ and $a_{i,\eta}$
is uniquely fixed in the non-Abelian gauge theory.
For simplicity, we replace the background fields by
$\tau$-independent and spatially constant
electric/magnetic fields, and consider SU(2) group.

\vspace*{1mm}
{\it Constant magnetic field} : 
Setting the constant magnetic field $B^z$ 
to the third color direction,
we use the gauge, $\alpha=0$ and 
$\alpha_i^a = \delta^{a3} (B/2)(y \delta_{i1} - x \delta_{i2})$.
By inspection, pure $a_\eta$ fluctuation is found stable,
and now we study the pure $a_i$ fluctuations.
After some algebra with help of the Gauss law constraint,
one can derive for $a_+^{(\pm)}$ with the third color charge $(\pm)$ 
and positive spin $+$ \cite{FujiiItakura,Chang-Weiss}
\begin{eqnarray}
\frac{1}{\tau}\partial_\tau (\tau \partial_\tau \tilde a_+^{(\pm)})
+ \left( \frac{\nu^2}{\tau^2} \mp mg B \pm 2gB\right) \tilde a_+^{(\pm)}
+ \left(-\partial_\perp^2 +\frac{g^2B^2 r^2}{4}\right) \tilde a_+^{(\pm)}
=0\, ,
\end{eqnarray}
where $m$ and $\nu$, respectively, are the orbital angular momentum and
the momentum conjugate to the rapidity
(and a similar equation for negative spin $-$).
Note that the term $\pm 2 gB$ stems from the anomalous magnetic moment.
Replacing the $x_\perp$-dependent part with the 2D harmonic oscillator
wavefunction ({\it Landau levels}),
the latter two terms in Eq.~(10) are combined and 
yields $({\nu^2 \over \tau^2} -g B) \tilde a_+^{(\pm)}$ 
for the lowest mode. 
At  $\sqrt{gB}\,\tau_{\rm wait}= \nu$,
this ``spring constant'' becomes negative:
{\it Nielsen-Olesen instability}\cite{Iwazaki,FujiiItakura}.
The typical scale $\sqrt{gB} \sim Q_s$.
Interestingly, this means that the mode with $\nu$ is stable until
$\tau_{\rm wait}=\nu/Q_s$, and then it grows up within 
$\tau_{\rm grow}=1/Q_s$. Conversely,
the maximum value $\nu_{\rm max}$ for the
unstable modes at $\tau$ satisfies 
$Q_s \tau = \nu_{\rm max} +1 \sim \nu_{\rm max}$.
This may be related
to the linear $\tau$-dependence of $\nu_{\rm max}$ found
in \cite{Raju}.

\begin{figure}[t]
\includegraphics[scale=0.6]{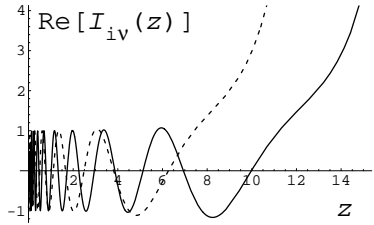}
\caption{
Re[$I_{i\nu}(z)$] for
$\nu$=8 (dashed) and 12 (solid), normalized at $z$=0.1.
The lowest mode of Eq.~(10) is given with the modified Bessel
function $I_{i\nu}(\sqrt{gB}\, \tau)$.
}
\end{figure}

\vspace*{1mm}
{\it Constant electric field} : 
Setting the constant electric field $E^z$ to the third
color direction, we take the gauge,
$\alpha^a=-(1/2)E\delta^{a3}$ and $\alpha_i =0$.
In the case of $a_\eta \neq 0$ and $a_i=0$, 
$a_\eta$ is found stable.
Let us consider the opposite case: $a_\eta=0$ and
$a_i \neq 0\, .$ Then the YM equations reduce to
\begin{equation}
{1 \over \tau} \partial_\tau  
( \tau \partial_\tau \tilde b^{(\pm)} )
+ \left\{
k_\perp^2 
+\frac{1}{\tau^2}
\left(\nu \pm \frac{ gE\tau^2}{2} \right)^2
\right\}\tilde b^{(\pm)} =0\, ,
\end{equation}
where $\tilde b(\tau, \nu, k_\perp)$ is related to
the transverse part of $\tilde a_i$. One sees that the spring constant
is non-negative and stable: no exponentially growing mode.
From the physics point of view, 
we expect, and are to clarify,
a certain relation of our formulation
to the Schwinger mechanism for the particle pair creation.

\section{Summary}
We have studied the time-evolution of the boost-invariant
color flux tube in a simplified Abelian-like setup, and
found the
outward expansion of the tube whose energy density evolves
similarly to the numerical simulations\cite{Lappi-McLerran}.
In the stability analysis we have 
found the unstable mode on the magnetic
background but no apparent unstable mode in the electric
case\cite{FujiiItakura,Iwazaki}.
To be more realistic,
we need to treat the time-dependent background field,
interactions among multi flux tubes, 
back-reactions of the fluctuations
on the background field, and so on.

\vspace{5mm}

%%%%%%%%%%%%%%%%%%%%%%%%%%%%%%%%%%%%%%%%%%%%%%%%%%%%%%%%%%%%

\end{document}